\documentclass[%
superscriptaddress,
showpacs,preprintnumbers,
 amsmath,amssymb,
 aps,twocolumn,
sort&compress,nofootinbib]{revtex4-1}

\usepackage[utf8]{inputenc}  
\usepackage[T1]{fontenc}
\usepackage{upgreek}

\usepackage{comment}
\usepackage{graphicx}
\usepackage{bm}
\usepackage{enumitem}
\usepackage{color}
\usepackage{multirow}
\usepackage[modulo,switch]{lineno}
\usepackage{adjustbox}	
\usepackage{placeins}	
\usepackage{dcolumn}
\newcolumntype{d}[1]{D{.}{.}{#1}}   

\newcommand{\tu}{\affiliation{Technische Universität Dresden, 01069 Dresden, Germany}}
\newcommand{\hzdr}{\affiliation{Helmholtz-Zentrum Dresden-Rossendorf (HZDR), 01328 Dresden, Germany}}
\newcommand{\ml}{\affiliation{INFN Sezione di Milano and Università degli Studi di Milano, 20122 Milan, Italy}}

\newcommand{\fin}{\affiliation{University of Jyväskylä, Department of Physics, P.O. Box 35, 40014 Jyväskylä, Finland}}

\begin{document}

\title{Measurement of the $^{2}$H($p,\gamma$)$^{3}$He S-factor at 265\,-\,1094\,keV}

\author{S.~Turkat}%
\tu

\author{S.~Hammer}%
\hzdr
\tu

\author{E.~Masha}%
\ml

\author{S.~Akhmadaliev}%
\hzdr

\author{D.~Bemmerer}%
\email{d.bemmerer@hzdr.de}%
\hzdr

\author{M.~Grieger}%
\hzdr
\tu

\author{T. Hensel}
\hzdr
\tu

\author{J. Julin}
\hzdr
\fin

\author{M. Koppitz}
\hzdr
\tu

\author{F. Ludwig}
\hzdr
\tu

\author{C. Möckel}
\hzdr
\tu

\author{S.~Reinicke}%
\hzdr
\tu

\author{R.~Schwengner}%
\hzdr

\author{K.~Stöckel}%
\hzdr
\tu

\author{T.~Szücs}%
\hzdr

\author{L.~Wagner}%
\hzdr
\tu

\author{K.~Zuber}%
\tu

\date{\today}
\begin{abstract}

Recent astronomical data have provided the primordial deuterium abundance with percent precision. As a result, Big Bang nucleosynthesis may provide a constraint on the universal baryon to photon ratio that is as precise as, but independent from, analyses of the cosmic microwave background. However, such a constraint requires that the nuclear reaction rates governing the production and destruction of primordial deuterium are sufficiently well known.

Here, a new measurement of the $^2$H($p,\gamma$)$^3$He cross section is reported. This nuclear reaction dominates the error on the predicted Big Bang deuterium abundance. A proton beam of 400\,-\,1650\,keV beam energy was incident on solid titanium deuteride targets, and the emitted $\gamma$-rays were detected in two high-purity germanium detectors at  angles of 55$^\circ$ and 90$^\circ$, respectively. The deuterium content of the targets has been obtained {\it in situ} by the $^2$H($^3$He,$p$)$^4$He reaction and offline using the Elastic Recoil Detection method. 

The astrophysical S-factor has been determined at center of mass energies between 265 and 1094 keV, addressing the uppermost part of the relevant energy range for Big Bang nucleosynthesis and complementary to ongoing work at lower energies. The new data support a higher S-factor at Big Bang temperatures than previously assumed, reducing the predicted deuterium abundance.

\end{abstract}


\maketitle
\section{\label{sec:intro}Introduction}

Recent measurements on the deuterium isotopic abundance in a "precision sample" of pristine clouds backlighted by quasars have led to a spectroscopic determination of the deuterium abundance from Big Bang nucleosynthesis, $N(^2{\rm H})/N({\rm H}) = (2.527\pm0.030) \times 10^{-5}$, i.e. a precision of 1.2\% \cite{Cooke18-ApJ}. These observational data offer a chance to determine the cosmological baryon density $\Omega_b$ independently from the cosmic microwave background (CMB). 

The most recent CMB-based value for $\Omega_b$ multiplied by the reduced Hubble parameter $h$ is found by the PLANCK 2018 data release to be $\Omega_b h^2\,=\,0.02236\pm0.00015$, resulting in a precision of 0.7\% \cite{Planck18_CosmologicalParameters}. When also lensing and baryon acoustic oscillation constraints are considered, a very similar value of $\Omega_bh^2\,=\,0.02242\pm0.00014$ is found \cite{Planck18_CosmologicalParameters}.

To good approximation, $\Omega_bh^2$ and the primordial deuterium abundance are linked by 
\begin{equation}
\frac{N(^2{\rm H})}{N({\rm H})} \propto \left[  \Omega_bh^2 \right]^{x}
\end{equation}
with an exponent $x$ ranging from  $x=$1.65 \cite{Pitrou18-PR} to $x=$1.598 \cite{Cyburt16-RMP}. 
As a result, the 1.2\% precise $N(^2{\rm H})/N({\rm H})$ value could in principle be converted to a 0.7\% precise constraint on $\Omega_bh^2$, which would be completely independent from CMB. Such a conversion presupposes that the nuclear physics of Big Bang deuterium production and destruction is precisely known. 

However, this is not the case. Detailed Big Bang nucleosynthesis (BBN) calculations with several different codes show that the final deuterium abundance after the completion of BBN is mainly influenced by deuterium destruction due to the $^2$H($d,p$)$^3$H, $^2$H($d,n$)$^3$He, and $^2$H($p,\gamma$)$^3$He reactions \cite{Serpico04-JCAP,Cyburt16-RMP,Pitrou18-PR,Moeckel19-Master}. Whereas the first two of these three reactions are relatively well known \cite{Greife95-ZPA,Leonard06-PRC,Huke08-PRC,Li17-PRC}, this is not true for the third one, $^2$H($p,\gamma$)$^3$He. 

\begin{figure*}[tb]
\includegraphics[width=\textwidth,clip]{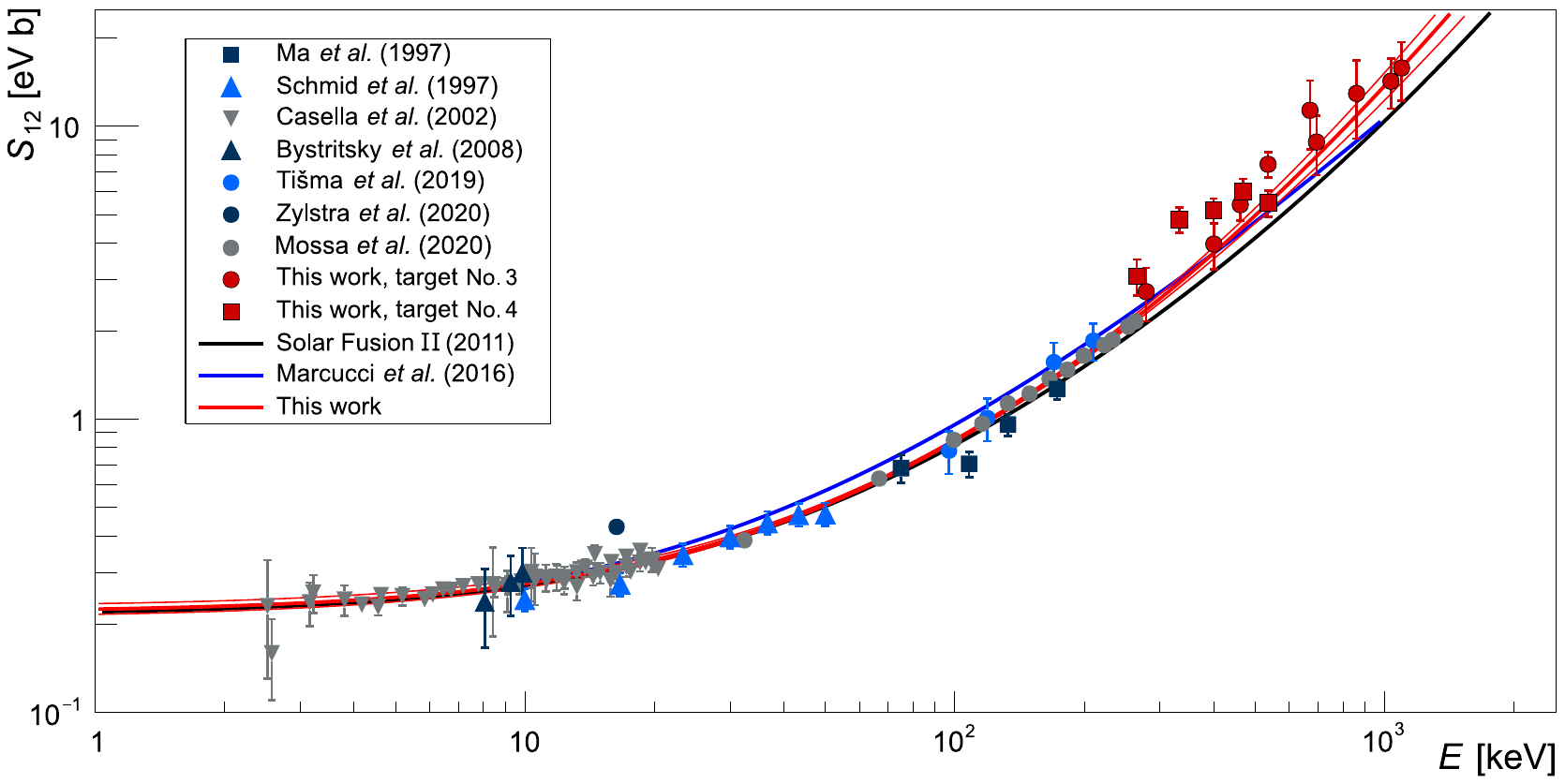}
\caption{\label{fig:sfactor}
Astrophysical S-factor of the $^2$H($p,\gamma$)$^3$He reaction as a function of the center of mass energy $E$. The symbols correspond to the experimental data of studies past 1990 by \cite{Schmid97-PRC, Ma97-PRC, Casella02-NPA, Bystritsky08-NIMA, Tisma19-EPJA, Zylstra20-PRC,Mossa20-Nature}. The Solar Fusion II \cite{Adelberger11-RMP} fit curve is shown in black. The theoretical \textit{ab initio} S-factor by Ref. \cite{Marcucci16-PRL} is shown in blue.}
\end{figure*}

The $^2$H($p,\gamma$)$^3$He reaction has been studied previously in experiments using a proton beam irradiating frozen heavy water targets \cite{Griffiths62-CJP,Griffiths63-CJP,Schmid97-PRC,Ma97-PRC,Bystritsky08-NIMA}, a windowless gas target deep underground \cite{Casella02-NPA,Mossa20-Nature}, or deuterated titanium targets \cite{Tisma19-EPJA}. Recently, $^2$H($p,\gamma$)$^3$He data have also been reported from an experiment using an inertially confined plasma of hydrogen and deuterium \cite{Zylstra20-PRC}. These data, expressed as the astrophysical S-factor \cite{Iliadis15-Book}
\begin{equation}
S_{12}(E) = \sigma(E) E \exp \left( \frac{25.64}{\sqrt{E / {\rm keV}}} \right)
\label{eq:sfactor}
\end{equation}
have been plotted in Figure \ref{fig:sfactor}, together with a parameterized fit from the 'Solar Fusion II' (SFII) community \cite{Adelberger11-RMP}.

Several of these experiments are confined to the low-energy\footnote{In this work, $E$ denotes the effective energy in the center of mass system, $E_p$ the proton beam energy in the laboratory system, and $E_\gamma$ the $\gamma$-ray energy.} region, $E<$\,25\,keV \cite{Schmid97-PRC,Bystritsky08-NIMA,Casella02-NPA,Zylstra20-PRC}, which corresponds to stellar and protostellar deuterium burning. 

The BBN energy range of interest is at higher energies, $E$\,=\,26\,-\,369\,keV for 95\% coverage, and has also been studied experimentally \cite{Griffiths62-CJP,Griffiths63-CJP,Ma97-PRC,Tisma19-EPJA,Mossa20-Nature}. Recently, there has been renewed interest in this energy region because of theoretical work using an {\it ab initio} approach resulting in an S-factor that is about 10\% higher than the SFII curve \cite{Marcucci16-PRL}. In this context, high-precision new data over most of the BBN range have very recently been reported from the LUNA underground ion accelerator in Gran Sasso \cite{Mossa20-EPJA,Mossa20-Nature}. 

The aim of the present work is to complement the new LUNA data \cite{Mossa20-Nature} with higher-energy cross section data, covering the upper part of the BBN energy region and, for normalization purposes, extending to even higher energies.

This work is organized as follows. The experimental setup and procedures are described in section \ref{sec:Experiment}. Section \ref{sec:tarana} is devoted to the determination of the deuterium content of the targets used. The S-factor data are derived and presented in section \ref{sec:results}. A parameterized fit of the data from the present work and from the literature is developed in section \ref{sec:sfactorfit}. Section \ref{sec:Discussion} discusses the data and fits. Finally, section \ref{sec:summary} includes a summary and outlook. Additional information can be found in thesis work \cite{Hammer19-Master,Masha19-Master}.

\begin{figure*}[tb]
\includegraphics[width=\columnwidth]{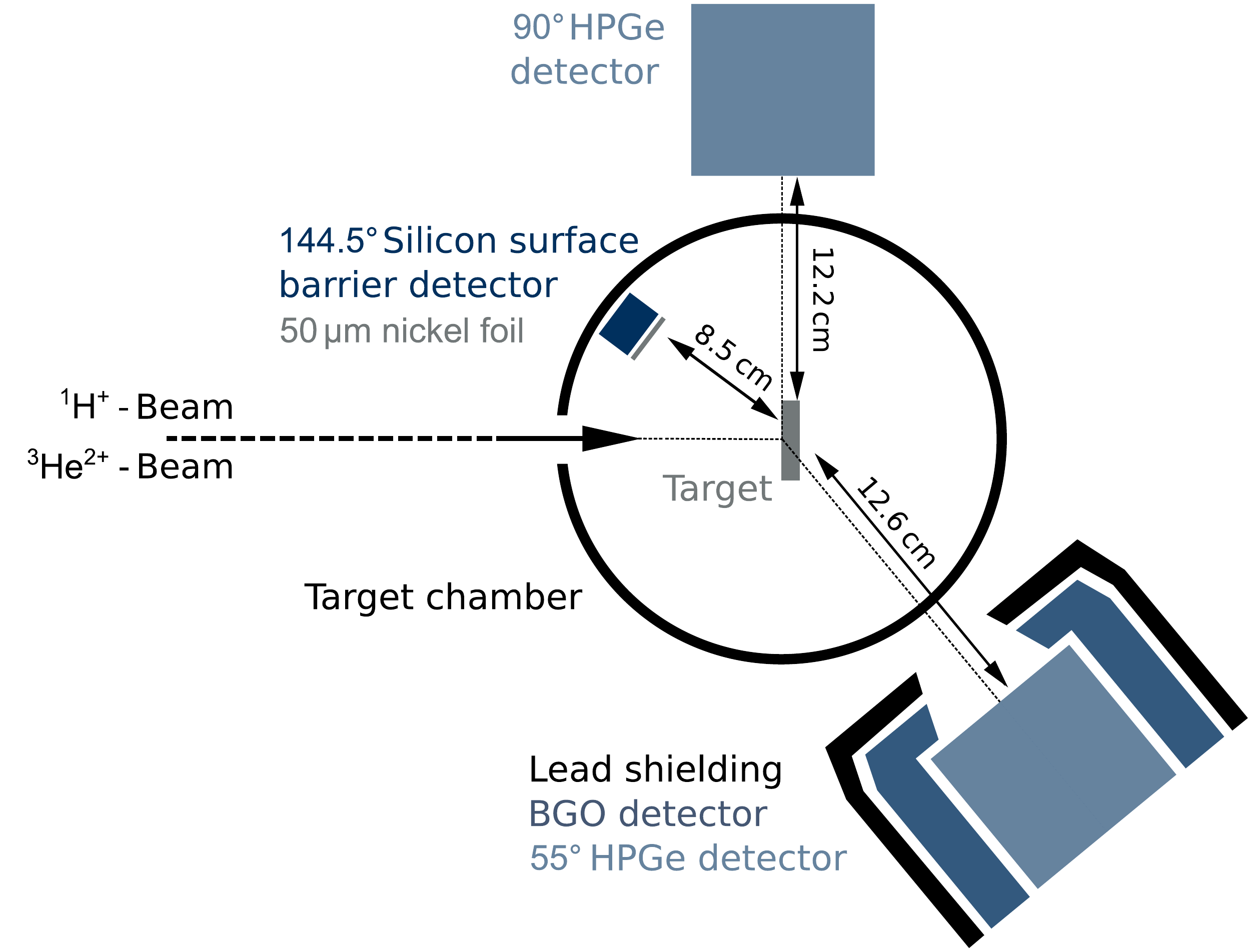}%
\includegraphics[width=\columnwidth]{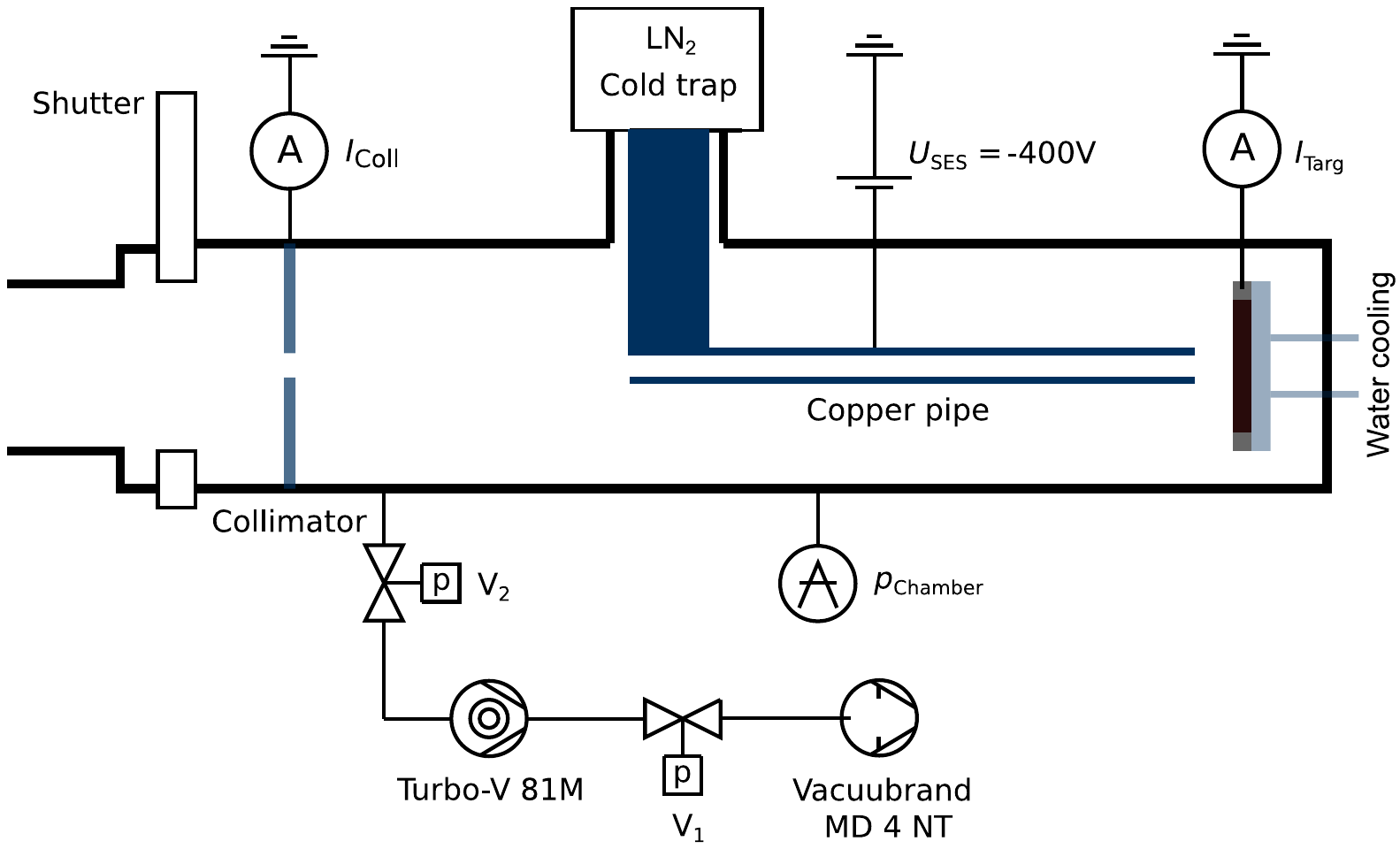}%
\caption{\label{fig:setup}\label{fig:targetchamber}%
Schematic view of the experimental setup with the beam entering from the left side. Left panel: Experimental setup showing the target chamber, the two HPGe detectors and the silicon detector seen from the top. Right panel: Setup including vacuum components and electronic devices seen from the side. For clarity, the copper pipe is shown in the right panel but omitted in the left panel. See text for details.}
\end{figure*}

\section{\label{sec:Experiment}Experiment}

The experiment has been conducted at the 3\,MV Tandetron accelerator in the Ion Beam Center \cite{Friedrich96-NIMA} of Helmholtz-Zentrum Dresden-Rossendorf, Dresden, Germany. 

\subsection{\label{subsec:accelerator}Accelerator and beam line}

For the main experiment, the 3\,MV Tandetron accelerator provided a $^1$H$^{+}$ beam with an energy $E_p$\,=\,400\,-\,1650\,keV and a beam intensity of $I$\,=\,3\,-\,4\,$\upmu$A on target, using a sputter ion source of type 860-C \cite{Friedrich96-NIMA}. 

For the {\it in situ} measurement of the deuterium areal density of the targets by the $^2$H($^3$He,$p$)$^4$He reaction, a Toroidal Volume Ion Source (TORVIS \cite{Hauser06-NIMB}) fed with a mixture of $^4$He and $^3$He gas was connected to the 3\,MV Tandetron, and a $^3$He$^{2+}$ beam of 1987\,keV energy and 30\,-\,40\,nA intensity was incident on the target.

Before reaching the target chamber, the accelerated ions passed a 11$^\circ$ switching magnet, horizontal and vertical electrostatic deflectors and a 7$^\circ$ misaligned beam line functioning as neutral particle trap.

\subsection{\label{subsec:chamber}Target chamber}

At the entrance of the target chamber (Figure \ref{fig:setup}), the beam passed a circular collimator of 5\,mm diameter, which absorbed 30-50\% of the total beam intensity, thus ensuring that the target was hit only by the homogeneous central part of the beam. A 12\,cm long copper pipe with 2.2\,cm outer and 1.8\,cm inner diameter was placed between the collimator and the target. This pipe was cooled to liquid nitrogen temperature in order to improve the local vacuum near the target and biased to -400\,V for the suppression of secondary electrons. Care was taken to avoid that the beam accidentally hits this tube. 

The targets were electrically insulated from the chamber and directly water cooled on their backings. The beam intensity was measured with an Ortec 439 digital current integrator, with an estimated precision of 1\% given by imperfections in the Faraday cup formed by the target and the copper pipe. The chamber was evacuated by an 80\,l/s turbomolecular pump. The pressure in the target chamber was always in the (0.6-6)\,$\times\,10^{-7}$\,hPa, equivalent to (0.4-4)\,$\times\,10^{-7}$ Torr, range. The good vacuum values reached confirm that there is no significant scattering of the incident beam off the liquid nitrogen cooled copper pipe. 

The target chamber wall consisted of 3\,mm stainless steel. Directly behind the 0.22\,mm thick tantalum target foil, there was a 30\,mm thick water column that was separated by a 2\,mm thick stainless stell wall from the target chamber vacuum. The target holder included up to 10\,mm of metal on the sides. This material may absorb $\gamma$-rays traveling from the target to the 90$^\circ$ detector, but it is not in the line of sight between the target and the 55$^\circ$ detector. In order to properly correct for the effects of these passive materials, the $\gamma$-ray detection efficiency was determined {\it in situ} using calibrated $\gamma$-activity standards and nuclear reactions, see section \ref{subsec:efficiency} below.

\subsection{\label{subsec:target} Targets}

Two different solid titanium deuteride targets, hereafter called targets No.\,3 and No.\,4, were used, each on a 220\,$\upmu$m thick tantalum backing with a diameter of 27\,mm. Tantalum had been selected because of its excellent mechanical characteristics and due to of the fact that according to local experience \cite{Marta10-PRC,Schmidt13-PRC,Schmidt14-PRC,Depalo15-PRC,Reinhardt16-NIMB,Wagner18-PRC} it does not give rise to significant ion beam induced $\gamma$-ray background. An important exception is the $^{19}$F($p,\alpha\gamma$)$^{16}$O reaction, which however produces well-identifiable $\gamma$-peaks that can be separated during the offline analysis.

Before the production of the targets, the backings were cleaned using a supersonic bath and solvent. Subsequently, a nominally 100\,nm thick layer of titanium was evaporated on them. Subsequently, the samples were deuterated by placing them in a deuterium gas atmosphere (nominal isotopic purity 99.99\%) at atmospheric pressure, slowly heated up (5 $^\circ$C/min) to 350\,$^\circ$C, kept at high temperature for 90 minutes, then again slowly cooled down to room temperature. 

\subsection{\label{subsec:detectors} Radiation and particle detectors}

The prompt $\gamma$-rays from the $^{2}$H($p,\gamma$)$^{3}$He reaction were detected using two high purity germanium  (HPGe) detectors. 

The first HPGe detector, with a relative efficiency\footnote{Full energy detection efficiency of the 1.33\,MeV $\gamma$-ray relative to a 3''$\times$3'' sodium iodide crystal, measured with a $^{60}$Co source placed at 25\,cm distance to the endcap on the symmetry axis of the crystal \cite{Gilmore08-Book}.} of 90\% and a HPGe crystal of 80\,mm diameter and  78\,mm length, was placed at an angle of 55$^\circ$ with respect to the beam axis. This detector was surrounded by a bismuth germanate (BGO) escape-suppression shield. The BGO was used to gate out events where the original photon does not deposit its complete energy in the HPGe crystal, but produces secondary photons which are escaping the detector. This may be due to Compton scattering or $e^+e^-$ pair production in the HPGe detector. The BGO shield is 5\,cm thick and extends to 7\,cm behind the back end of the crystal. The frontal part of the BGO also has a thickness of 5\,cm and a circular entrance hole of 6\,cm diameter. An additional layer of lead (3\,cm) acts as a passive shielding around the BGO and 7\,cm of lead covered the BGO on the front side. It is noted that the BGO shield also reduces the continuum background from cosmic-ray muons in the HPGe detector \cite{Szucs19-EPJA}. 

The second $\gamma$-ray detector, with 60\% relative efficiency and a HPGe crystal of 71\,mm diameter and  60\,mm length, was placed at an angle of 90$^\circ$ with respect to the beam axis, without any active or passive shielding. It served mainly to estimate possible effects of the $\gamma$-ray angular distribution on the final data. 

For the {\it in situ} target thickness measurement, a partially depleted silicon surface barrier detector with an active area of 300\,mm$^2$ and a sensitive thickness of 2000\,$\upmu$m was used. The detector was shielded from elastically scattered beam particles by a 50\,$\upmu$m thick nickel foil, which was placed in front of the detector. While elastically scattered beam particles were stopped in the nickel foil (the range of 4\,MeV protons in nickel is approximately 50\,$\upmu$m), the foil is transparent to the high-energy protons from the $^2$H($^3$He,$p$)$^4$He reaction ($Q\,=\,18.353\,$MeV).

\subsection{\label{subsec:DAQ}Data acquisition}

The data from the two HPGe detectors, the BGO guard detector, and the Si detector were fed to separate channels of a 16-channel, 14-bit, 250 MS/s digitiser of type CAEN V1725. The digitiser was operated in pulse height analysis mode with a trapezoidal filter algorithm, using the firmware provided by the manufacturer. 

The ion beam currents measured on the target and on the final collimator were digitised using Ortec 439 digital current integrators, and the digital outputs were fed into additional channels of the V1725 digitiser. All channels were individually self-triggered.

The digitiser and its parameters were controlled by the CAEN MC2 software, and the data were timestamped and recorded in list mode on the computer hard disk for offline analysis. During the offline analysis, the timestamped events were sorted and grouped in time. The inclusion of the current integrator made it possible to exclude significant variations in ion beam current and yield during each irradiation. 

For quick online analysis, a second, independent data acquisition system was maintained, using Ortec 671 main amplifiers and a histogramming Ortec 919E unit. The final analysis, however, was always performed on the V1725 list mode data.

\subsection{\label{subsec:irradiations}Irradiations}

For the irradiations, the cross section was studied in 14 different runs at energies ranging from $E_p$ = 403-1644 keV. For the sake of consistency, the runs at $E_p$ = 807\,keV and 605\,keV were repeated for both targets No.\,3 and No.\,4. As a result, altogether 12 different beam energies between $E_p$\,=\,403\,-\,1644\,keV have been studied.

After mounting each target, the chamber was evacuated for several hours until a vacuum below  $10^{-6}$\,hPa (10$^{-6}$ Torr)  was reached, and then the irradiation was started. Before dismounting a target, the copper pipe was brought to room temperature, and the chamber was vented with dry nitrogen from a bottle.

Depending on the beam energy and the status of the target, running times of 1.4\,-\,10.1\,h and integrated charges of 30\,-\,137\,mC were used.

\section{\label{sec:tarana} Target analysis}

At the proton beam energies used here, $E_p$ = 403-1644 keV, the beam loses 9-21\,keV of energy when passing the target, leading to a 0.5-4\% decrease of the cross section over the entire target thickness. Therefore, for the purposes of the present analysis it is sufficient to determine the total deuterium content, integrated over the entire target thickness, and correct for this small decrease in cross section in the offline analysis.

The deuterium content of the targets was determined using two different methods: The elastic recoil detection (ERD) method was applied to both targets (section \ref{subsec:erda}). In addition, target No.\,4 was studied {\it in situ} using the nuclear reaction analysis (NRA) method (section \ref{subsec:nra}).

It is noted that for both of these methods of target analysis, due to the uniform production process (section \ref{subsec:target}) it is assumed that the targets are homogeneous in the direction perpendicular to the beam.

\subsection{\label{subsec:erda} Elastic Recoil Detection Analysis}

The ERD study was performed after the conclusion of the main experiment. For this purpose, the targets were removed from the setup and brought to the ERD setup of the 6\,MV Tandetron accelerator of the HZDR ion beam center.

There, each sample was bombarded with a 43\,MeV $^{35}$Cl$^{7+}$ beam of low intensity (typically 200-350 pA) at a grazing angle of 75$^\circ$ between beam axis and sample normal. Two representative spots of 1.5\,mm $\times$ 1.5\,mm were chosen for the analysis, one inside the visible proton beam spot (hereafter called {\it on spot}), and the second in an area of the target that was several mm away and not touched by the proton beam (hereafter called {\it off spot}). 

In order to exclude that the {\it off spot} position was affected by temperature dependent effects such as outgassing, heat transport simulations were carried out using the Energy2D Interactive Heat Transfer Simulation software \cite{Xie12-PhysicsTeacher} package. The  simulations showed that heat conductance in the 0.22\,mm thick tantalum target backing to the cooling water was so efficient that for positions more than 1\,mm outside the beam spot, the temperature rise was less than 5\,K. Therefore, thermal effects can be neglected for the  {\it off spot} position. 

The elastically recoiled target atoms were detected in a Bragg ionization chamber that was installed at an angle of 31$^\circ$ with respect to the beam.

\begin{figure}[tb]
\includegraphics[width=\columnwidth]{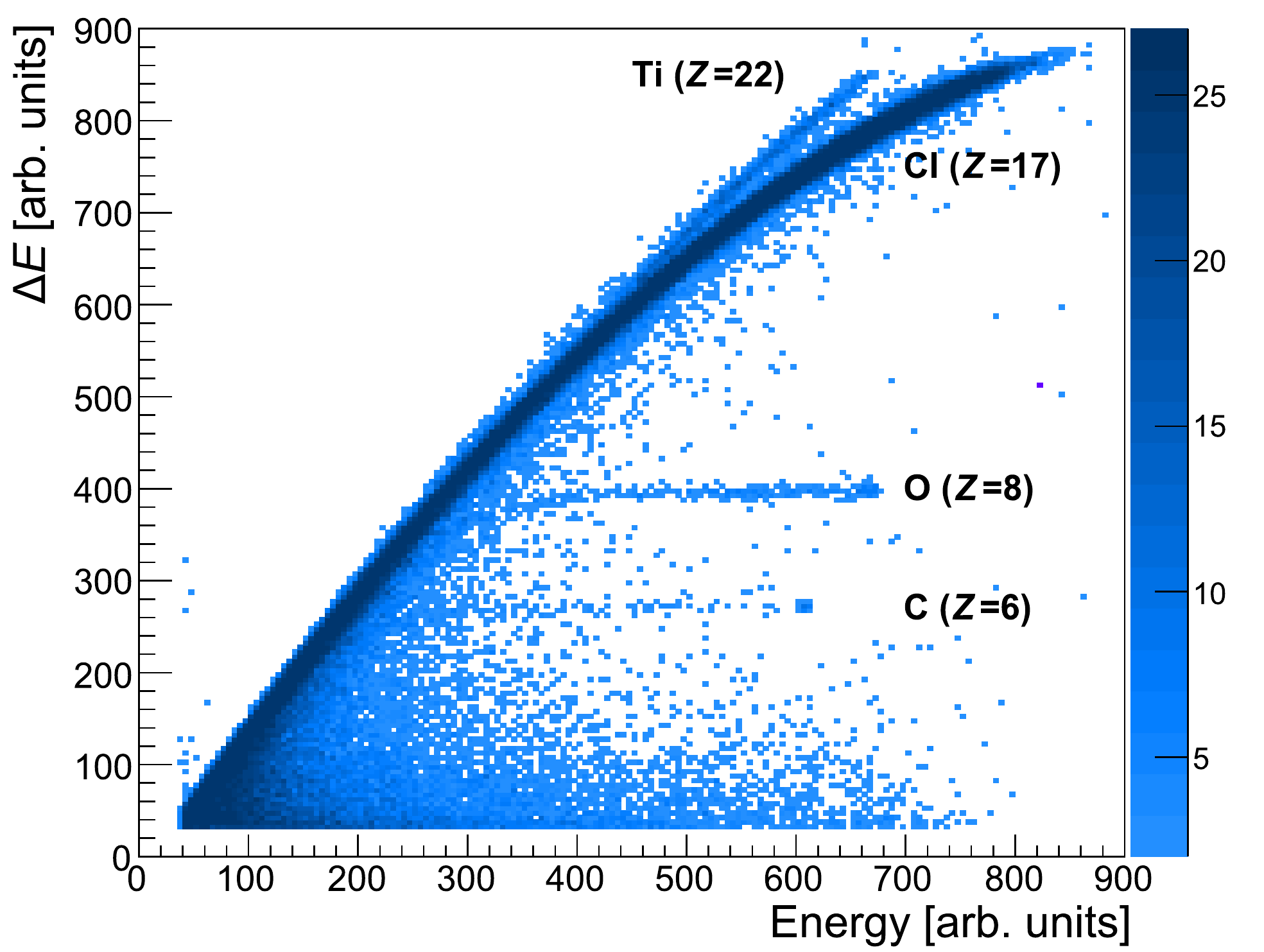}
\caption{\label{fig:ERDAheavy}$\Delta E-E$ histogram of the elastic recoil detection analysis with the Bragg chamber, at the center of the visible beam spot of target No.\,4. See text for details.}
\end{figure}

The two-dimensional histogram by the Bragg chamber (Figure \ref{fig:ERDAheavy}) showed elastically scattered chlorine from the beam, and a branch due to the titanium target material. In addition, significant amounts of oxygen and carbon were found in both targets, much of it in the initial layer. Finally, the spectra shows a hint of possible nitrogen.

\begin{figure}[tb]
\includegraphics[width=\columnwidth]{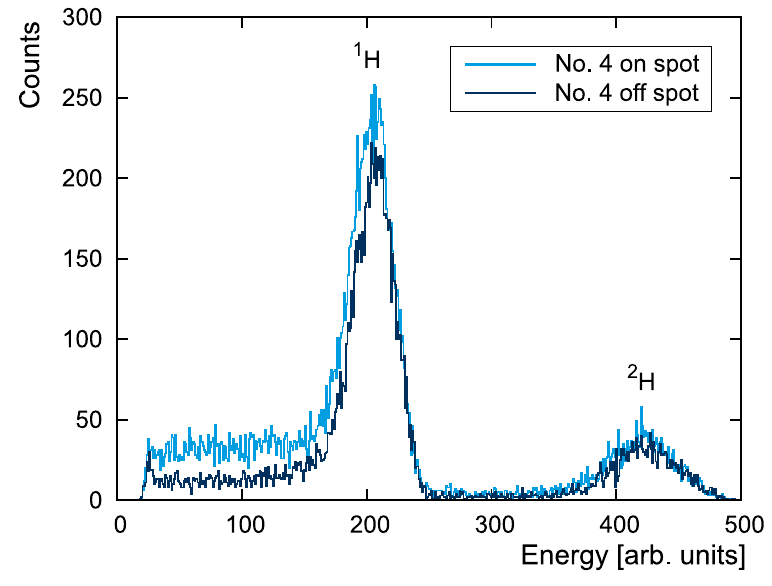}
\caption{\label{fig:ERDAlight}Pulse height spectra of the silicon detector used for determining the $^{1,2}$H content in the ERD runs, for target No.\,4, inside and outside the respective proton beam spot. The two runs have similar incident charge but the spectra were not normalized.}
\end{figure}

The Bragg chamber does not resolve the light recoils hydrogen and deuterium. They were studied instead using a silicon particle detector at an angle of 41$^\circ$ with respect to the beam. Typical pulse height spectra from the silicon detector are shown in Figure \ref{fig:ERDAlight} for target No.\,4. 

The target composition was then derived from the ERD data using the NDF v9.3g software \cite{Barradas97-APL}. 
Taking both hydrogen isotopes $^{1,2}$H together, the approximate stoichiometry in the main layer of the target is TiH$_{0.6}$, well below the aimed stoichiometry of TiH$_2$ (titanium hydride). 
It is noted that both samples studied showed some roughness, so that the ERD depth profiles are affected by an increased uncertainty. For the cross section determination, only the integrated deuterium thickness from ERD is needed, which should be unaffected.

The high amount of $^1$H, which is comparable to $^2$H in all cases, may in part be explained by pre-existing hydrogen in the tantalum backing that may have migrated from the Ta to the Ti during the deuteration process.
In addition, the chamber used for the deuteration of the titanium targets is usually operated with natural hydrogen, and it cannot be fully excluded that due to dead volumes in the gas system some residual hydrogen was present during the deuteration process.
Also, it is possible that during storage and transport before or after the main experiment water vapor from ambient air may have modified the target.
Finally, the observed increase of $^1$H content {\it on spot} with respect to the {\it off spot} case (Figure \ref{fig:ERDAlight}) is consistent with $^1$H$^+$ ions from the ion beam that are stopped in the backing.

The resulting deuterium areal density in the targets ranges from (267-293)$\times$10$^{15}$ cm$^{-2}$ (Table \ref{tab:ERDAtab}). In order to be conservative, the unweighted average of the two ERD runs was used for each target. The statistical error bar was then enlarged to cover the values of the areal densities derived at the two different positions studied on a given target, resulting in 6-8\% statistical error for the $^2$H areal density (Table \ref{tab:ERDAtab}). In this way, possible changes in the integrated deuterium areal density during the irradiation are included in the 1$\sigma$ error bars.

An additional 10\% systematic (scale-factor) uncertainty was estimated to account for uncertainties in the calibration of the ERD apparatus.

\subsection{\label{subsec:nra}Nuclear reaction analysis}

\begin{figure*}[!htb]
\includegraphics[width=\textwidth]{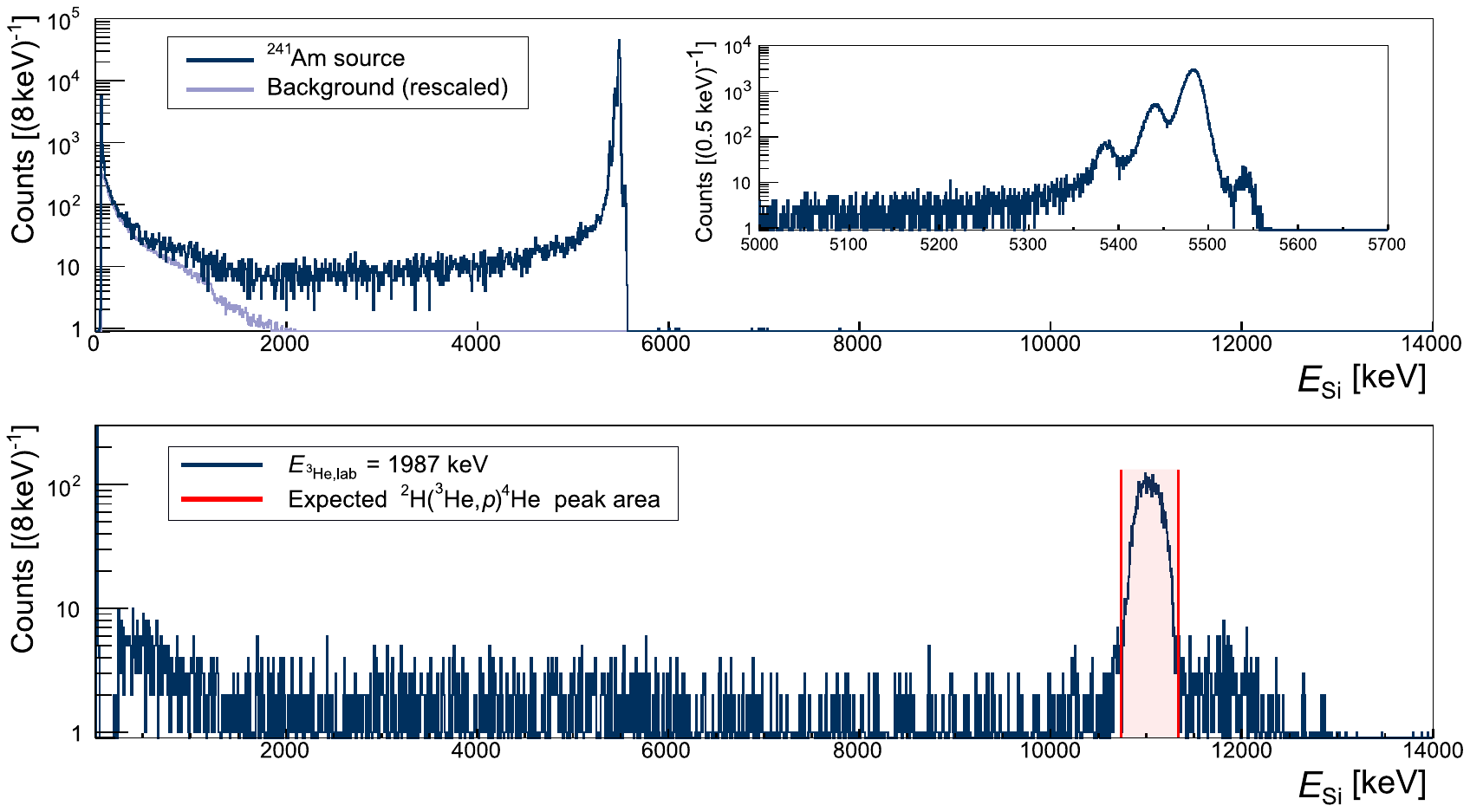}
\caption{\label{fig:particlespec}Top: Pulse height spectra in the {\it in situ} silicon particle detector of a $^{241}$Am source (blue; spectrum taken without the 50\,$\upmu$m nickel foil) and the no-source background (grey). The inset shows the region of interest with a smaller binning.
--- Bottom: Pulse height spectrum from the irradiation of the deuterated titanium sample No.\,4 with a $^3$He$^{2+}$ beam. The expected energy range for protons from the $^2$H($^3$He,$p$)$^4$He reaction, after passing the 50\,$\upmu$m nickel foil, is given by red lines.
}
\end{figure*}

\begin{table}[tb]
\caption{Hydrogen areal densities of the targets used in units of 10$^{15}$\,cm$^{-2}$, as obtained from the ERD and NRA analyses. For the ERD $^2$H results, the first error bar is statistical, the second the 10\% ERD scale factor uncertainty. See text for details. }
\centering
\begin{tabular}{c|c|c|c|c} 
\hline
\hline
Method & \multicolumn{2}{c|}{ERDA} & NRA & Adopted\\ 
Target &  $^1$H  &  $^2$H  & $^2$H & $^2$H\\
\hline
No.\,3 &282 $\pm$ 16& 293 $\pm$ 23 $\pm$ 29	&  &   293 $\pm$ 23\\
No.\,4 &311 $\pm$ 21& 267 $\pm$ 17 $\pm$ 27 & 285 $\pm$ 15 & 282 $\pm$ 14\\
\hline
\hline
\end{tabular}
\label{tab:ERDAtab}
\end{table}

For the nuclear reaction analysis (NRA), the $^2$H($^3$He,$p$)$^4$He reaction ($Q$\,=\,18.353 MeV) was used. Its cross section has been studied previously \cite[and references therein]{Wielunska16-NIMB}. The particle detector to detect the protons from the reaction is placed in backwards geometry, at an angle of 144.5$^\circ$ in the laboratory frame with respect to the beam direction. 

A $^3$He beam energy of $E_{^3\rm He}$ = 1987 keV has been selected. There is a wide resonance in the $^2$H($^3$He,$p$)$^4$He reaction at lower energies, $E_{^3\rm He} \approx$ 600\,keV that was avoided here. At $E_{^3\rm He}$ = 1987 keV, the $^3$He beam loses $\Delta E_{^3\rm He}$ = 65 keV in the deuterated layer of the target. Over this energy range, the cross section increases by 2.5\%, leading to an effective cross section of (12.9$\pm$0.6)\,mb/sr \cite{Wielunska16-NIMB}. 

For the NRA run, a TORVIS-type ion source \cite{Hauser06-NIMB} was fed with a special mixture of $^3$He (5\% by volume) and $^4$He gas (95\% by volume) and connected to the 3\,MV Tandetron accelerator. Using this gas mixture, a tuneable beam intensity of typically 1\,$\upmu$A He$^-$ was extracted from the ion source. After magnetic analysis by the injection magnet before the tandem, the $^3$He$^-$ current was 30\,nA (at the tandem entrance) and the $^3$He$^{2+}$ current on the target was limited to just 7-12\,nA (3.5-6 particle-nA), in order to avoid overloading detector and data acquisition system. 

The peak due to the $^2$H($^3$He,$p$)$^4$He reaction shows up at exactly the predicted energies (Figure \ref{fig:particlespec}), after taking into account energy loss and straggling in the protective nickel foil. At the right of the main peak, there is a second peak due to the parasitic $^{14}$N($^3$He,$p$)$^{16}$O reaction ($Q$\,=\,15.243\,MeV) on small nitrogen impurities in the target bulk. Two further $^2$H($^3$He,$p$)$^4$He runs, at $^3$He beam energies of 1987 and 2188\,keV, had to be discarded because for these two runs, the parasitic peak grew over time to be similar in size to the main $^2$H($^3$He,$p$)$^4$He peak. It is possible that the nitrogen impurity grew over time, as nitrogen from the residual gas bonded to the target surface under the influence of the cold finger.

The solid angle and energy calibration of the silicon detector were determined using a calibrated $^{241}$Am activity standard with an $\alpha$ emission rate known to 2\% precision (95\% confidence level). For this measurement, the $^{241}$Am source was placed exactly at the target position. The 50\,$\upmu$m protective nickel foil was removed in order to be able to measure the $\alpha$ particles from the $^{241}$Am source. Despite the large active depth of 2000\,$\upmu$m, the particle detector displayed sufficient resolution to resolve all four $^{241}$Am peaks (Figure \ref{fig:particlespec}).

Due to beam time constraints and the delayed arrival of the $^3$He-$^4$He gas mixture, only target No.\,4 could be studied using the NRA method. After correcting for the $^3$He beam charge and silicon detector solid angle, the deuterium areal density due to NRA was (285$\pm$15)$\times10^{15}$ cm$^{-2}$, in good agreement with the ERD value of (267$\pm$17$_\text{stat})\times10^{15}$ cm$^{-2}$ for the same target (Table \ref{tab:ERDAtab}). As a consequence, for target No.\,4 the weighted average of the two methods is adopted for the further analysis (Table \ref{tab:ERDAtab}).

\section{\label{sec:results}Data Analysis and Results}

\begin{figure*}[t!]
\includegraphics[width=2.0\columnwidth]{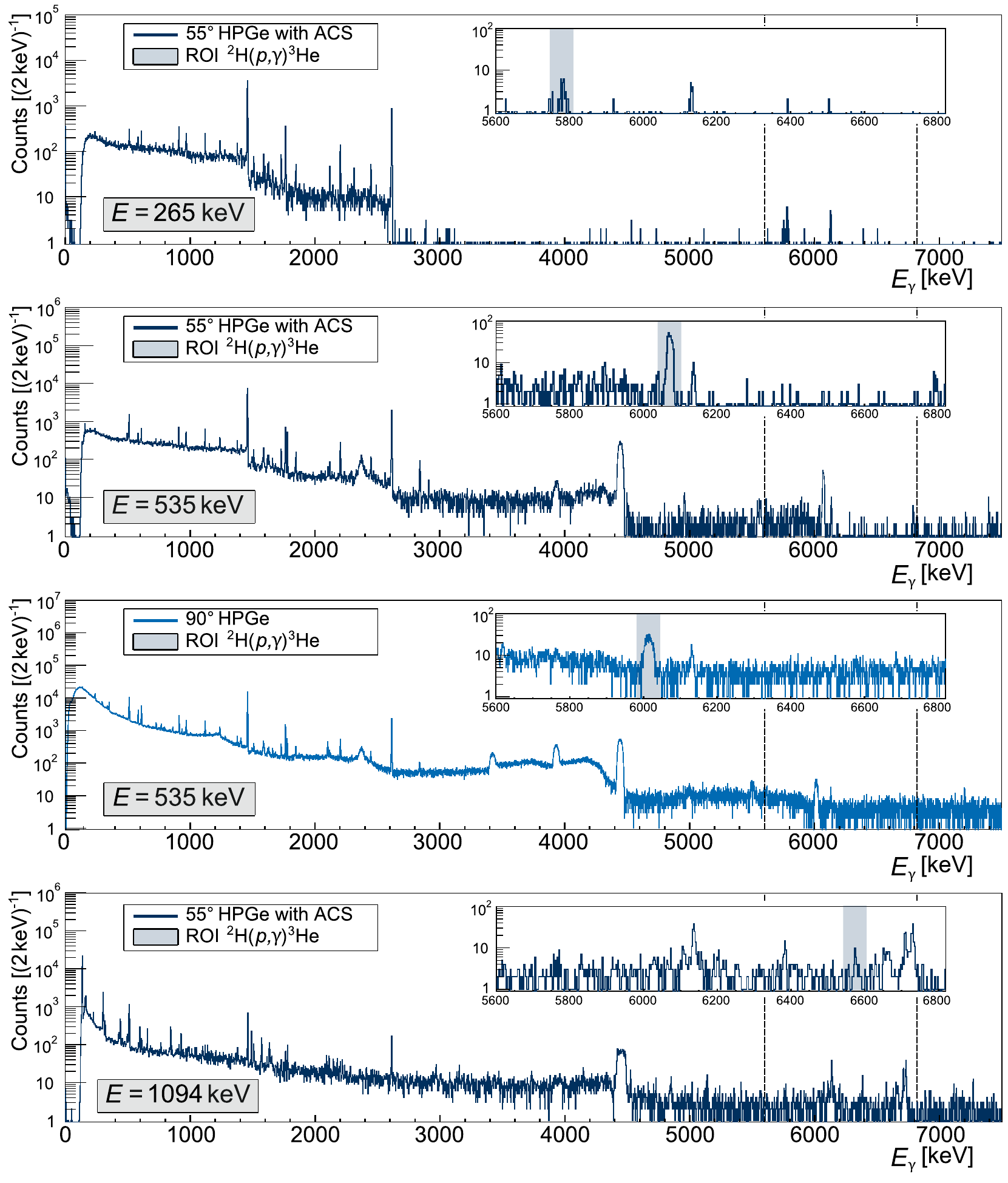}
\caption{\label{fig:inbeamspec} $\gamma$-ray spectra at the lowest and highest energies studied, $E$\,=\,265\,and\,1094\,keV, respectively, and at a representative intermediate energy $E$\,=\,535\,keV. The shaded areas inside the insets show the beam energy and detector angle dependent region of interest for the $^2$H($p,\gamma$)$^3$He peak. See text for details.}
\end{figure*}

In this section, the analysis of the $\gamma$-ray spectra and the determination of the astrophysical S-factor are described.

\subsection{Efficiency calibration of the $\gamma$-ray detectors}
\label{subsec:efficiency}

The energy and efficiency calibration of the detectors were performed using activity standards of $^{137}$Cs, $^{88}$Y, and $^{60}$Co provided by Physikalisch-Technische Bundesanstalt (Braunschweig, Germany) with typically 1\% activity uncertainty. These standards were point-like and provided enclosed between two 0.2\,mm thick polyethylene foils that were glued together by heating. 

Due to the significant amounts of passive materials included in the target chamber and target holder (section \ref{subsec:chamber} and Figure \ref{fig:targetchamber}), great care was taken to ensure that the detection efficiency was determined in exactly the same geometry as the one used for the irradiations. As a first step, the activity standards, which were embedded in thin foils,
were mounted atop a tantalum backing on the target holder. The water cooling was kept running during the determination of the $\gamma$-ray efficiency in order to also include $\gamma$-ray attenuation in the cooling water. 

The full-energy detection efficiency curve was then extended to high energies using the well-known $E_p$\,=\,991\,keV resonance in the $^{27}$Al($p$,$\gamma$)$^{28}$Si reaction \cite{Anttila77-NIM,Zijderhand90-NIMA}, applying the two-line method following a procedure used before \cite{Marta10-PRC,Schmidt13-PRC,Schmidt14-PRC,Depalo15-PRC,Reinhardt16-NIMB,Wagner18-PRC}. The resulting full energy $\gamma$-ray detection efficiency in the relevant energy range of $E_\gamma$\,=\,5800\,-\,6600\,keV was between 7.7-6.7\,$\times\,10^{-4}$ for the $90^{\circ}$ detector and between 2.6-2.3\,$\times\,10^{-4}$ for the $55^{\circ}$ detector. The final uncertainty of the detection efficiency was  3\%, taking into account the branching ratios in $^{27}$Al($p$,$\gamma$)$^{28}$Si, the statistics of the $^{27}$Al($p$,$\gamma$)$^{28}$Si and offline runs, the calibration of the activity standards, and the fit uncertainty of the efficiency curve. 

\subsection{Interpretation of the $\gamma$-ray spectra}

For all proton beam energies $E_p$, the low $\gamma$-ray energy part of the spectrum, $E_\gamma$ $<$ 3\,MeV, is dominated not by beam-induced effects but by the laboratory background. The laboratory background is given by $^{40}$K ($E_\gamma$\,=\,1461\,keV) and the natural decay chains, most prominently by $^{208}$Tl decay ($E_\gamma$\,=\,2615\,keV). 
The weak 1779\,keV line is likely caused by the $^{27}$Al($p,\gamma$)$^{28}$Si reaction on the target holder. It has been used to place an experimental upper limit of 0.5\% on beam protons that are elastically scattered off the copper tube and hit the aluminum target holder.

For the lowest proton beam energy studied here, $E_p$\,=\,403\,keV ($E$\,=\,265\,keV), in addition to the signal from the reaction under study, there is a weak background due to the $^{19}$F($p$,$\alpha\gamma$)$^{16}$O reaction at 6130\,keV. Fluorine is a well-known contaminant in tantalum target backings, and the peak is well separated from the region of interest (Figure \ref{fig:inbeamspec}, top panel).

At a typical intermediate proton beam energy,  $E_p$\,=\,807\,keV ($E$\,=\,535\,keV), again the $^{19}$F($p$,$\alpha\gamma$)$^{16}$O line is evident but still sufficiently well separated from the region of interest. In addition, there is now a strong background due to the 4439\,keV line by the decay of the first excited state of $^{12}$C populated in the $^{15}$N($p$,$\alpha\gamma$)$^{12}$C reaction \cite{Imbriani12-PRC,Reinicke14-Diplom}, which is however well below the energy range of interest (Figure \ref{fig:inbeamspec}, second and third panels).

At the highest energy studied here, $E_p$\,=\,1644\,keV ($E$\,=\,1094\,keV), additional channels in the $^{19}$F($p$,$\alpha\gamma$)$^{16}$O background reaction are opening, giving rise to a number of peaks in the 6\,-\,7\,MeV $\gamma$-ray energy region (Figure \ref{fig:inbeamspec}, bottom panel). Still, it was possible to extract the signal from the reaction under study assuming a constant background fitted on sidebands in the $\gamma$-ray spectrum.

\begin{figure}[t!]
\includegraphics[width=\columnwidth]{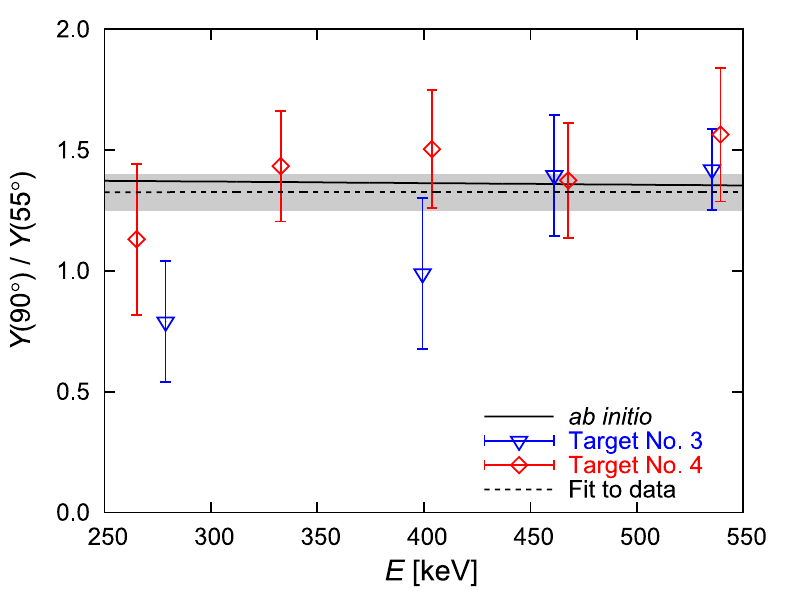}
\caption{\label{fig:Yieldratio} Ratio of the efficiency-corrected yields $Y^\ast$ of both HPGe detectors as a function of center-of-mass energy $E$. The {\it ab initio} predicted yield ratio is shown as a solid line \cite{Girlanda10-PRL,Marcucci16-PRL,Marcucci17-PC}. The dashed curve and shaded area are a constant fit to the data and its uncertainty. See text for details.
}
\end{figure}

\subsection{\label{subsec:angdist} Cross-check of the assumed angular distribution}

For the data analysis, the theoretical angular distribution of the prompt $\gamma$-rays from the deexcitation into the ground state of $^3$He determined by Marcucci and co-workers was used \cite{Girlanda10-PRL,Marcucci16-PRL,Marcucci17-PC}. In order to check this assumption, the ratio of efficiency-corrected yields for the two detectors used here was determined (Figure \ref{fig:Yieldratio}). This was only possible at low center-of-mass energies, $E$ = 265\,-\,535\,keV. 
At higher energies, the unshielded 90$^\circ$ detector was affected by too high background for a reliable analysis.   

The data show a 90$^\circ$/55$^\circ$ yield ratio that can be described by a constant value of 1.33$\pm$0.08,
in very good agreement with the {\it ab initio} predicted value of 1.38 which is almost independent of energy (Figure \ref{fig:Yieldratio}).

As a result, also for higher energies, where the 90$^\circ$ detector could not be analyzed due to beam induced background, the {\it ab initio}  $\gamma$-ray angular distribution \cite{Girlanda10-PRL,Marcucci16-PRL,Marcucci17-PC} is used, assuming a 6\% uncertainty obtained from the yield ratio fit (Figure \ref{fig:Yieldratio}).

\subsection{\label{subsec:sfactor}Determination of the astrophysical S-factor}

At each beam energy, the astrophysical S-factor given by eq.~(\ref{eq:sfactor}) was calculated for each detector separately, corrected for the assumed {\it ab initio} angular distribution (section \ref{subsec:angdist}). The data for both angles were in mutual agreement within their relative statistical uncertainties, and a weighted average was formed (Table \ref{tab:s-factor}).
The only exception is the $E$\,=\,279\,keV data point for target No.\,3. There, the angle-corrected S-factor values for the two detectors differ by 1.6 standard deviations.

\begin{figure}[t!]
\includegraphics[width=\columnwidth]{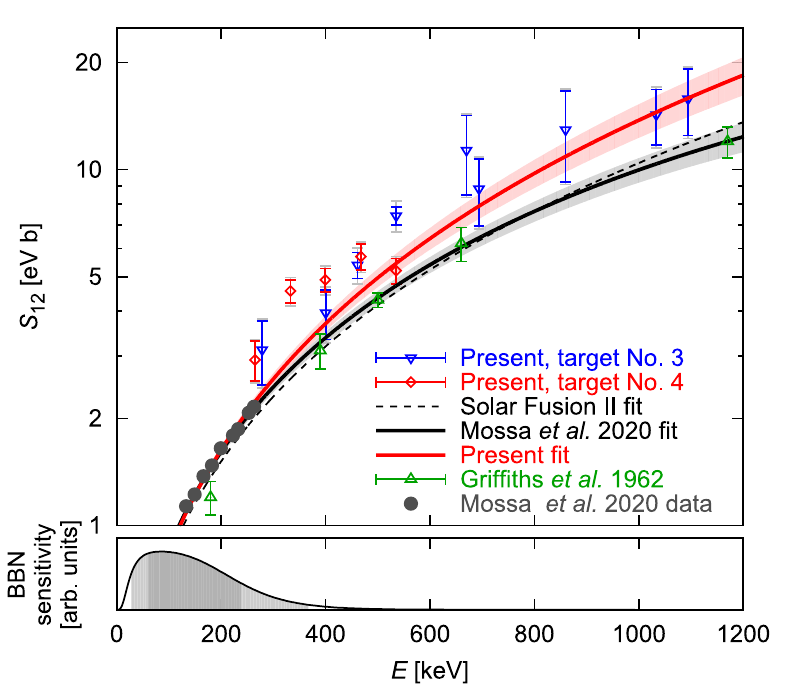}
\caption{\label{fig:SfactorTurkat} Experimental S-factor data from the present work for the two different targets used. The inner error bar is the statistical error, the outer error bar reflects statistical and systematic errors combined in quadrature. Previous fits  \cite{Adelberger11-RMP,Mossa20-Nature} and the present new fit are also shown  (shaded area for the error band \cite[present]{Mossa20-Nature}). Previous data in this energy range \cite{Griffiths62-CJP,Mossa20-Nature} are also included.
The lower panel shows the BBN sensitive energy range (light grey: 95\% of the area, dark grey: 68\% of the area). See text for details.}
\end{figure}

When comparing both targets, overall fair agreement is found (Figure \ref{fig:SfactorTurkat}). Furthermore, all the new data points lie above the previously recommended Solar Fusion II \cite{Adelberger11-RMP} curve.

\begin{table}[b]
\caption{\label{tab:s-factor} Experimental S-factors $S(55^\circ)$ and $S(90^\circ)$ of the $^{2}$H($p,\gamma$)$^{3}$He reaction from the analysis of both detectors, corrected for the theoretical $\gamma$-ray angular distribution. In addition, the weighted average S-factor of both detectors is listed, as well as the relative statistical and systematic uncertainties.
}
\begin{tabular}{crr*{2}{d{4}}d{2}rr}
\hline
\hline
Target & \multicolumn{1}{c}{$E_\textrm{p}$} & \multicolumn{1}{c}{$E$} & \multicolumn{1}{c}{$S(55^\circ)$} &\multicolumn{1}{c}{$S(90^\circ)$} & \multicolumn{1}{c}{$S$} & $\displaystyle\frac{\Delta S_\text{stat}}{S}$	&	$\displaystyle\frac{\Delta S_\text{syst}}{S}$\\ \cline{4-6} 
&  \multicolumn{1}{c}{[keV]} & \multicolumn{1}{c}{[keV]} & \multicolumn{3}{c}{[eV\,barn]} & \multicolumn{2}{c}{[\%]}\\
\hline \hline
No.\,3 & 424 & 279 &  4.7(11)  &2.7(6) &	3.1	& 21 &	8\\
 & 605 & 400 & 4.9(12)   &3.6(7) &	4.0    & 16 &	\\ 
 & 696 & 461 &  5.3(8)  &5.5(6) &	5.4	&	8 &	\\ 
 & 807 & 535 &  7.2(7)  &7.6(6) &	7.4	&	6 &	\\ 
 & 1009 & 670 & 11.3(29)   & &	11.3	&	25 &	\\ 
 & 1044 & 694 & 8.8(19)   & &	8.8	&	21 &	\\ 
 & 1293 & 860 & 13.(4)  & &	12.9	&	29 &	\\
 & 1553 & 1033 & 14.3(25)   & &	14.3	&	18 &  \\ 
 & 1644 & 1094 & 14.(5)   & 17.(5) &	15.8	&	21 &  \\ 
\hline
No.\,4 & 403 & 265 & 3.3(8)   & 2.8(5) &	2.9	&	13 &	5\\ 
 & 504 & 333 & 4.4(6)   & 4.7(5) &	4.6	&	8 &	\\ 
 & 604 & 400 & 4.6(6)   &   5.1(5) &	4.9 &	8 &	\\ 
 & 706 & 468 & 5.6(8)   &5.8(6) &	5.7	&	8 &	\\ 
 & 807 & 535 & 4.7(7)   &5.5(6) &	5.2	&	8 &	\\ 
\hline
\hline
\end{tabular}
\end{table}

\begin{figure*}[tb]
\includegraphics[width=\textwidth]{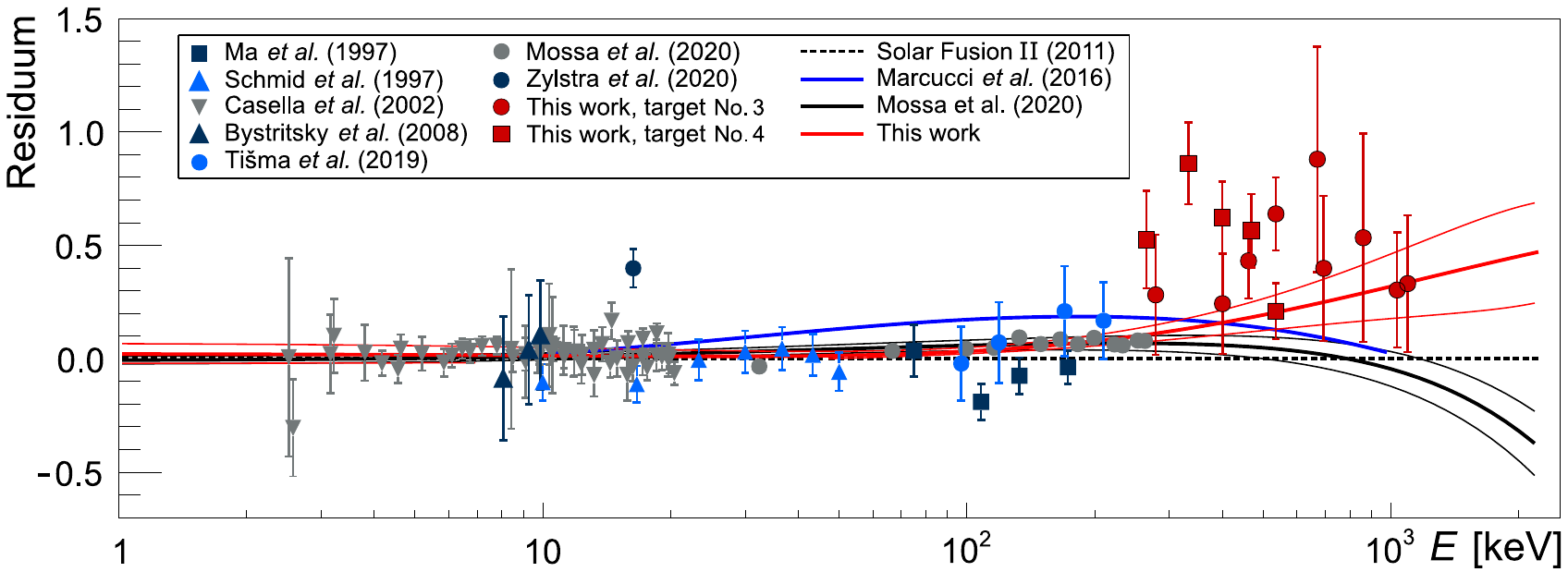}
\caption{\label{fig:Residuum} Residuum of the S-factor data \cite[and present work]{Schmid97-PRC, Ma97-PRC, Casella02-NPA, Bystritsky08-NIMA, Tisma19-EPJA, Zylstra20-PRC, Mossa20-Nature} with respect to the Solar Fusion II fit \cite{Adelberger11-RMP}. The {\it ab initio} theory curve by Marcucci \cite{Marcucci16-PRL}, the Mossa {\it et al.} fit \cite{Mossa20-Nature}, and the present fit are also shown (blue, black, and red, respectively). The standard deviation (1$\sigma$) of the present fit is given by thin red lines. The error bars for the data points are only statistical.}
\end{figure*}

\subsection{\label{subsec:uncertainties}Uncertainties}

The statistical uncertainties vary from point to point and are dominated by the counting statistics in the HPGe detectors, which contribute 6\,-\,29\% to the error budget (Table \ref{tab:uncertainty}). They are generally higher for higher beam energies, where the ion beam induced background played a larger role.

The systematic uncertainty is divided into two parts: First, a component that is common for a given target out of the two targets used but separate for the two targets. For target No.\,3, it amounts to 8\% and is given by the variation between the two ERDA beam spots studied (section \ref{subsec:erda}). For target No.\,4, it amounts to 5\% and is given by the weighted average of the ERDA uncertainty (12\%; 10\% from normalisation and 6\% point to point variation for target No.\,4) and the NRA uncertainty, which is 6\% (geometric sum of 4.6\% from the reference cross section \cite{Wielunska16-NIMB}, 3\% from the NRA run statistics, and 1\% from the solid angle determination). 

The second part of the systematic uncertainty is a 12\% scale factor uncertainty affecting all the data from the present work. It is composed of 10\% from the calibration of the ERDA apparatus and method (section \ref{subsec:erda}), 6\% from the angular correction (section \ref{subsec:angdist}), 3\% from the $\gamma$-ray detection efficiency, and 1\% from the beam current integration. It is noted that for target No.\,4, this adopted uncertainty is conservative, because there the ERDA scale factor uncertainty already plays a small role in the per-target error, and again in the global scaling error.

\begin{table}[b!]
\caption{\label{tab:uncertainty} Contributions to the statistical, systematic, and scale-factor uncertainties for the $^{2}\textrm{H}(p,\gamma)^{3}\textrm{He}$ S-factor. }
\begin{ruledtabular}
\begin{tabular}{lrrr}
Contribution & Stat. & Syst. & Scale \\
\hline
HPGe peak entries & 6-29\% &\\
ERDA beam spot variation & & 6-8\% \\ 
ERDA normalization & & & 10\% \\ 
NRA normalization & & 4.6\% & \\ 
NRA statistics & & 3.0\% & \\ 
Angular correction & & & 6\% \\ 
$\gamma$-ray detection efficiency & & & 3\% \\
Beam intensity & & & 1\%\\ \hline
Total & 6-29\% & 5-8\% & 12\% \\ 
\end{tabular}
\end{ruledtabular}
\end{table}

\section{\label{sec:sfactorfit}Parametrized fit of the S-factor curve}

For the purpose of the discussion, the parameterized fit of the S-factor curve $S_{12}(E)$ derived in the Solar Fusion II \cite{Adelberger11-RMP} compilation is repeated here, adding the present new data and also the post-Solar Fusion II data by Ti{\v{s}}ma {\it et al.} \cite{Tisma19-EPJA}, LUNA \cite{Mossa20-Nature}, and Zylstra {\it et al.} \cite{Zylstra20-PRC}. Pre-1990 work \cite{Griffiths62-CJP,Griffiths63-CJP,Bailey70-CJP} is left out of the fit due to the generally much less detailed description of the experiment, when compared to post-1990 work \cite{Casella02-NPA,Ma97-PRC,Schmid97-PRC,Bystritsky08-NIMA,Tisma19-EPJA,Mossa20-Nature,Zylstra20-PRC}. 

Following Solar Fusion II \cite{Adelberger11-RMP}, the shape of the S-factor curve is taken as
\begin{equation}
S_{12}(E) = S^{0}_{12} + S'_{12} \left(\frac{E}{\rm keV}\right) + S''_{12} \left(\frac{E}{\rm keV}\right)^2
\label{eq:Adelshapeeq}
\end{equation}
with the parameters $S^{0}_{12}$, $S'_{12}$, and $S''_{12}$ to be determined by the fit.

As a first step, Equation (\ref{eq:Adelshapeeq}) was fitted on all the post-1990  data \cite[and present work]{Casella02-NPA,Ma97-PRC,Schmid97-PRC,Bystritsky08-NIMA,Tisma19-EPJA,Mossa20-Nature,Zylstra20-PRC}, using a quadratic sum of statistical and systematic error bars for each data point. This initial fit served to determine the shape of the excitation function.

Subsequently, the fit was repeated, in turn, for each data set $i$  (with $i$ denoting each of the datasets \cite[and present work]{Casella02-NPA,Ma97-PRC,Schmid97-PRC,Bystritsky08-NIMA,Tisma19-EPJA,Mossa20-Nature,Zylstra20-PRC}) separately. During this second fit, the previously determined parameters $S^{0}_{12}$, $S'_{12}$, and $S''_{12}$ were kept fixed. For each data set $i$, a normalization factor $\alpha_i$ was then determined scaling the fixed S-factor shape:
\begin{eqnarray}\label{eq:Adelshape}
S^{i}_{12}(E) &=& \alpha_i \times S_{12}(E)  \\
    &=& \alpha_i \times \left[ S^{0}_{12} + S'_{12} \left(\frac{E}{\rm keV}\right) + S''_{12} \left(\frac{E}{\rm keV}\right)^2 \right] \nonumber
\end{eqnarray}
For References \cite{Casella02-NPA,Ma97-PRC,Schmid97-PRC,Bystritsky08-NIMA,Tisma19-EPJA,Zylstra20-PRC}, in the second step only the statistical uncertainties were used for each data point. For the present work, the ERDA beam spot variation error, which is not common to the entire data set, is added in quadrature. For the new LUNA data \cite{Mossa20-Nature}, the total errors are used. The numbers of data points $N_i$, resulting $\chi_i^2$ values and normalization factors $\alpha_i$ are listed in Table \ref{tab:analysistab}. In order to properly account for possibly discrepant data sets, following Ref. \cite{Adelberger11-RMP}, for each data set it is checked whether the condition 
\begin{equation}\label{eq:chisqcondition}
\chi_i^2 \leq \chi_i(0.5) \overset{!}{=} \chi^2(N_i-1;0.5)
\end{equation}
is fulfilled. Here, $\chi^2(N_i-1;0.5)$ is the value of the $\chi^2$ function for $N_i-1$ degrees of freedom for a probability value of $P=0.5$. 
If condition (\ref{eq:chisqcondition}) is fulfilled, the fitted $\alpha_i$ value and its error bar are kept unchanged for the next step. This applies to most of the data sets studied here \cite{Casella02-NPA,Schmid97-PRC,Bystritsky08-NIMA,Tisma19-EPJA,Mossa20-Nature}.

If instead, condition  (\ref{eq:chisqcondition}) is not fulfilled, $\Delta \alpha_i$ is scaled up by a factor $\sqrt{\chi_i^2/\chi_i(0.5)}$, again following the Solar Fusion II compilation \cite{Adelberger11-RMP}. This applies to Ref. \cite{Ma97-PRC} ($f_\text{Ma}$\,=\,1.15) and to the present work ($f_\text{Turkat}$\,=\,1.53).

\begin{table}[tb]
\caption{Results of fitting Equation (\ref{eq:Adelshape}) for the individual data sets. $N_i$ is the number of data points, $\chi_i^2$ the resulting $\chi^2$ from the fit, $\chi_i^2(0.5)$ the $\chi_i^2$ for 50\% probability, $f_i$ the inflation factor, $\alpha_i(\Delta \alpha_i)$ the scaling factor and its uncertainty before inflation, and $\varepsilon_i$ the relative systematic uncertainty of data set $i$. See text for details.}
\centering
\begin{tabular}{lrd{2}d{2}d{2}d{7}r} 
\hline
\hline
Reference &$N_i$& \multicolumn{1}{c}{$\chi_i^2$} &  \multicolumn{1}{c}{$\chi_i^2(0.5)$} & \multicolumn{1}{c}{$f_i$}  & \multicolumn{1}{c}{$\alpha_i$} &  \multicolumn{1}{c}{$\varepsilon_i$} \\
\hline
Casella {\it et al.} \cite{Casella02-NPA} & 51 & 27.0 & 49.3 &1    & 1.008(9) & 4.5\% \\
Ma {\it et al.} \cite{Ma97-PRC} & 4 & 3.2 &2.4&1.15& 0.88(4) & 9\% \\
Schmid {\it et al.} \cite{Schmid97-PRC} &7 &3.5&5.3 &1  &  0.96(3)  & 9\% \\
Bystritsky {\it et al.} \cite{Bystritsky08-NIMA} & 3 & 0.28 &1.4 &   1  &  1.01(14) & 8\% \\
 Ti{\v{s}}ma {\it et al.} \cite{Tisma19-EPJA}  &4 & 0.7 &2.4    &   1 &  1.04(8) & 10\% \\
 Zylstra {\it et al.} \cite{Zylstra20-PRC}  &1 & && 1    &  1.38(8) & 17\% \\
 Mossa {\it et al.} \cite{Mossa20-Nature}  &13 &8.5 &11.3& 1    &  0.999(8) & 2.6\% \\
Present work &14 & 29.0 &12.3&1.53&  1.22(3)  & 12\% \\ \hline
Global average & & & &  & 1.000(21) \\ 
\hline
\hline
\end{tabular}
\label{tab:analysistab}
\end{table}

The very recent laser based work by Zylstra {\it et al.} \cite{Zylstra20-PRC} has just one data point but lies somewhat above the fit curve. Still, no inflation factor is used for Zylstra {\it et al.}, considering also that, when including the 17\% systematic uncertainty by that work, the data are close to the SFII curve.

Using the data of Table \ref{tab:analysistab}, with an uncertainty of $\sqrt{(f_i\Delta\alpha_i)^2+(\alpha_i\varepsilon_i)}$ for each data set $i$, a weighted average of $\bar{\alpha}\,=\,1.000\,\pm\,0.021$ is found. If the outlier data by Zylstra {\it et al.} \cite{Zylstra20-PRC} were to be omitted, the global average would reduce to $0.998\pm0.021$. 

The final results for the parameterization of the S-factor curve using Equation (\ref{eq:Adelshape}) are
%
\begin{eqnarray}\label{eq:Sfactorfitresults}
S_{12}^0 &=& (0.219 \pm 0.004) \, \textrm{eV\,b} \nonumber \\
S_{12}' &=& (5.4 \pm 0.2) \times 10^{-3} \, \textrm{eV\,b} \\
S_{12}'' &=& (8.1 \pm 0.9) \times 10^{-6} \, \textrm{eV\,b}  \nonumber 
\end{eqnarray}
and are included in Figures \ref{fig:sfactor}, \ref{fig:SfactorTurkat}, and \ref{fig:Residuum}. 

Finally, the energy dependent impact of the $^2$H($p,\gamma)^3$He S-factor on the BBN $^2$H/H abundance predictions was studied using the following relation
\begin{equation}
S_{12}^{\text{mod}}(E,E_k) = S_{12}(E) \left[1+C\times\delta(E-E_k)\right]
\label{eq:sfacpar}
\end{equation}
where $C$ denotes an arbitrary constant and $S_{12}(E)$ is given by Equation (\ref{eq:Sfactorfitresults}). 
 The $E_k$ were chosen as integer values in the range of 0\,-\,2000\,keV. The resulting sensitivity is plotted in the lower panel of Figure \ref{fig:SfactorTurkat}. 
 
 It is found that the present data lie just above the inner sensitive range for Big Bang nucleosynthesis, but need to be included in order to reach 95\% coverage, required for precision conclusions (Figure \ref{fig:SfactorTurkat}).

\section{\label{sec:Discussion}Discussion }

The present data points from the two targets studied here are generally in mutual agreement, but there are two outliers for target No.\,4, at $E$\,=\,333 and 535\,keV. The relevant runs for these data points were done directly after $^3$He$^{2+}$ beam runs for NRA analysis, and it is in principle possible that the necessary complete re-optimization of the beam parameters after the $^3$He run may have led the beam to be focused on other parts of the target than those covered by the ERD analysis. The resultant increased scatter of the present data around the overall fit curve results in an inflation factor of $f_\text{Turkat}$\,=\,1.53 (section \ref{sec:sfactorfit}) that is taken into account in the analysis. 

When leaving the two outliers aside, the present data points show mutual agreement and are consistently higher than the previous Solar Fusion II fit (Figure \ref{fig:Residuum}). They show a slope to higher energies that is increasingly steeper than Solar Fusion II. Also, the present data do not confirm the slight downward turn seen in the high-energy part of the energy dependence of the {\it ab initio} curve \cite{Marcucci16-PRL}, possibly due to that work focusing mostly on lower energies  \cite{Marcucci16-PRL}. Indeed, at $E$\,=\,250\,-\,550\,keV, the {\it ab initio} predicted $\gamma$-ray angular distribution is in good agreement with the present data. 

The present data are significantly higher than the only previous data points in the same energy region, by Griffiths and co-workers \cite{Griffiths62-CJP}, see Figure \ref{fig:SfactorTurkat}. In that work, both ice and gas targets were used, and the absolute cross sections were determined based on a yield measurement at 90$^\circ$ angle. The $\gamma$-ray angular distribution used for the analysis by Griffiths {\it et al.} gives about 10\% higher values at 90$^\circ$ than indicated by the present work, explaining part of the discrepancy. Another effect may be a possible oversubtraction of $^2$H($d,n$)$^3$He-induced background caused by elastically scattered deuterium in the work by Griffiths {\it et al.} with their very thick ice targets \cite{Griffiths62-CJP}. For the present thin targets, $^2$H($d,n$)$^3$He background plays no role.

At lower energies, $E$\,=\,10\,-\,200\,keV, the present S-factor fit curve closely follows the previous Solar Fusion II fit (Figure \ref{fig:Residuum}). In this energy range, which corresponds directly to the most important range for BBN (Figure \ref{fig:SfactorTurkat}), there are no data from the present work, but there are the new high-precision gas target data from LUNA \cite{Mossa20-EPJA,Stoeckel20-PhD}. Within errors, the present data are not far from the new LUNA data points at similar energies. 

However, there are some differences between the present fit and the LUNA fit \cite{Mossa20-Nature}. The two fits are virtually indistinguishable, $\leq$1.2\% difference, for $E \leq$ 200\,keV, due to the strong effect of the high-precision LUNA data, which are included in both fits. However, at $E$ = 400 keV, where the present new data have been added, the present fit is 16\% higher than Solar Fusion II \cite{Adelberger11-RMP} and 10\% higher than LUNA  \cite{Mossa20-Nature}. 

All available post-1990 data are well described by the present fit curve. The only exception is the laser-based data point by Zylstra \cite{Zylstra20-PRC}, which used a novel technique that is not yet widely adopted for nuclear cross section measurements. It is noted that at the low energy studied \cite{Zylstra20-PRC}, a 10\% error in the effective temperature of the laser implosion capsule would lead to 22\% change in the determined S-factor. 

\section{\label{sec:summary}Summary and outlook}

The $^2$H($p,\gamma$)$^3$He cross section has been studied using solid deuterated titanium targets in the center of mass energy range $E$\,=\,265\,-\,1094\,keV, starting from the upper end of the BBN relevant energy window and extending to higher energy. The target thickness has been determined by the elastic recoil detection method and, for one of the two targets, by nuclear reaction analysis using the well-known $^2$H($^3$He,$p$)$^4$He reaction.

The data show that previous S-factor fits \cite{Adelberger11-RMP,Mossa20-Nature} may have to be revised upwards, in the energy range studied here.

\begin{acknowledgments}

The authors are indebted to Oliver Busse (TU Dresden) for kindly deuterating the titanium samples, to Laura Marcucci (Pisa) for kindly providing the \textit{ab initio} cross sections and angular distributions, to Andreas Hartmann (HZDR) for technical support, and to Bernd Scheumann and Claudia Neisser (HZDR) for preparing the evaporated samples. --- 
Financial support by Deutsche Forschungsgemeinschaft DFG (ZU123/21-1, BE4100/4-1), Konrad-Adenauer-Stiftung, the Helmholtz Association (ERC-RA 0016), and the ChETEC COST Action CA16117 is gratefully acknowledged.

\end{acknowledgments}



\end{document}